# Locating the Source of Diffusion in Large-Scale Networks

Pedro C. Pinto,[1] Patrick Thiran,[1] Martin Vetterli[1]

[1]École Polytechnique Fédérale de Lausanne (EPFL), Lausanne, Switzerland

How can we localize the source of diffusion in a complex network? Due to the tremendous size of many real networks—such as the Internet or the human social graph—it is usually infeasible to observe the state of all nodes in a network. We show that it is fundamentally possible to estimate the location of the source from measurements collected by sparsely-placed observers. We present a strategy that is optimal for arbitrary trees, achieving maximum probability of correct localization. We describe efficient implementations with complexity $O(N^\alpha)$, where $\alpha = 1$ for arbitrary trees, and $\alpha = 3$ for arbitrary graphs. In the context of several case studies, we determine how localization accuracy is affected by various system parameters, including the structure of the network, the density of observers, and the number of observed cascades.

Localizing the source of a contaminant or a virus is an extremely desirable but challenging task. In nature, many animals are intrinsically capable of performing source localization. Through chemotaxis, for example, certain bacteria can analyze concentration gradients around them in order to quickly move towards the source of a nutrient, or quickly avoid the source of a poison [1,2]. Animals such as the Pacific salmon and the green sea turtles are capable of using olfaction to navigate in odor plumes, for foraging or reproductive activities [3,4]. In certain systems, however, the task of localizing the source has to be performed in a *network*, rather than in the continuous space. This is the case, for example, when an infectious disease spreads through human populations across a large region, as observed with the worldwide H1N1 virus pandemic in 2009. Here the system is more conveniently modelled as a network of interconnected people, and source localization reduces to identifying which person in the network was first infected.

In recent years, there has been significant effort in studying the dynamics of epidemic outbreaks on networks [5–11]. In particular, the focus has been on the *forward problem* of epidemics: understanding the diffusion process and its dependence on the rates of infection and cure, as well as on the structure of the network. In this letter, we focus on the *inverse problem* of inferring the original source of diffusion, given the infection data gathered at some of the nodes in the network. The ability to estimate the source is invaluable in helping authorities contain the epidemic or contamination. In this context, the inference of the underlying propagation network was studied in [12], while the inference of the unknown source was analyzed in [13], in both cases assuming that we know the state of *all nodes* in the network. More recently, the controllability of complex networks was considered in [14], using appropriately selected driver nodes. Here, our goal is to locate the source of diffusion under the practical constraint that only a *small fraction of nodes* can be observed. This is the case, for example, when locating a spammer who is sending undesired emails over the Internet, where it is clearly impossible to monitor all the nodes. Thus, the main difficulty is to develop tractable estimators that can be efficiently implemented (i.e., with sub-exponential complexity), and that perform well on multiple topologies.

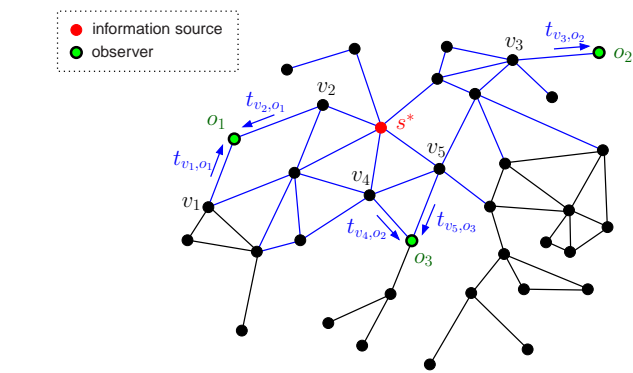

Figure 1. Source estimation on an arbitrary graph $\mathcal{G}$. At the unknown time $t = t^*$, the information source $s^*$ initiates the diffusion. The blue edges denote those over which information has already propagated. In this example, there are three observers, which measure *from which neighbours* and *at what time* they received the information. The goal is to estimate, from these observations, which node in $\mathcal{G}$ is the information source.

We first introduce our network model. The underlying network on which diffusion takes place is modeled by a finite, undirected graph $\mathcal{G} = \{V, E\}$, where the vertex set $V$ has $N$ nodes, and the edge set $E$ has $L$ edges (Fig. 1). The graph $\mathcal{G}$ is assumed to be known, as is often verified in practice—e.g., rumors spreading in a social network, or electrical perturbations propagating on the electrical grid. The *information source*, $s^* \in \mathcal{G}$, is the vertex that originates the information and initiates the diffusion. We model $s^*$ as a random variable (RV) whose prior distribution is uniform over the set $V$, i.e., any node in the network is equally likely to be the source a priori.

The diffusion process is modeled as follows. At time $t$, each vertex $u \in \mathcal{G}$ has one of two possible states: i) *informed*, if it has already received the information from any neighbour; or ii) *ignorant*, if it has not been informed so far. Let $\mathcal{V}(u)$ denote the set of vertices directly connected to $u$, i.e., the neighbourhood or vicinity of $u$. Suppose $u$ is in the *ignorant* state and, at time $t_u$, receives the information for the first time from one neighbour—say, $s$—thus becoming *informed*. Then, $u$ will retransmit the information to all its other neighbours, so that each neighbour $v \in \mathcal{V}(u) \backslash s$ receives the information at time $t_u + \theta_{uv}$, where $\theta_{uv}$ denotes the random *propagation delay* associated with edge $uv$. The RVs $\{\theta_{uv}\}$ for different edges $uv$ have a known, arbitrary joint distribution. The diffusion process is initiated by the source $s^*$ at an unknown time $t = t^*$. This diffusion model is general enough to accommodate various scenarios encountered in practice.

Let $O \triangleq \{o_i\}_{i=1}^{K} \subset \mathcal{G}$ denote the set of $K$ observers, whose location on $\mathcal{G}$ is chosen or known. Each observer measures *from which neighbour* and *at what time* it received the information. Specifically, if $t_{v,o}$ denotes the absolute time at which observer $o$ receives the information from its neighbour $v$, then the *observation set* is composed of tuples of direction and time measurements, i.e., $\mathcal{O} \triangleq \{(o, v, t_{v,o})\}$, for all $o \in O$ and $v \in \mathcal{V}(o)$.

How is the source location recovered from the measurements taken at the observers? We adopt a *maximum probability of localization* criterion, which corresponds to designing an estimator $\hat{s}(\cdot)$ such that the localization probability $P_{\text{loc}} \triangleq \mathbf{P}(\hat{s}(\mathcal{O}) = s^*)$ is maximized. Since we consider $s^*$ to be uniformly random over $\mathcal{G}$, the optimal estimator is the maximum likelihood (ML) estimator,

$$
\begin{aligned}
\hat{s}(\mathcal{O}) &= \operatorname*{argmax}_{s \in \mathcal{G}} \mathbf{P}(\mathcal{O}|s^* = s) \\
&= \operatorname*{argmax}_{s \in \mathcal{G}} \sum_{\Pi_s} \mathbf{P}(\Pi_s|s^* = s) \times \\
&\qquad \int \cdots \int g(\theta_1, \cdots, \theta_L, \mathcal{O}, \Pi_s) d\theta_1 \cdots d\theta_L.
\end{aligned} \quad (1)
$$

Here, $\Pi_s$ is the set of *all possible paths* $\{\mathcal{P}_{s,o_k}\}_{k=1}^{K}$ between the source $s$ and the observers in the graph $\mathcal{G}$; the set $\{\theta_l\}_{l=1}^{L}$ represents the random propagation delays for all $L$ edges of graph $\mathcal{G}$; and $g$ is a deterministic function that depends on the joint distribution of the propagation delays in a complicated way. In essence, the estimator in (1) is performing averages over two different sources of randomness: a) the uncertainty in the paths that the information takes to reach the observers, and b) the uncertainty in the time that the information takes to cross the edges of $\mathcal{G}$. Due the combinatorial nature of (1), its complexity increases exponentially with the number of nodes in $\mathcal{G}$, and is therefore intractable. In what follows, we propose a strategy of complexity $O(N)$ that is optimal for general trees, and a strategy of complexity $O(N^3)$ that is sub-optimal for general graphs.

Consider first the case of an underlying tree $\mathcal{T}$. Because a tree does not contain cycles, only a subset $O_{\text{a}} \subseteq O$ of the observers will receive information emitted by the unknown source. We call $O_{\text{a}} = \{o_k\}_{k=1}^{K_{\text{a}}}$ the set of $K_{\text{a}}$ *active observers*. The observations made by the nodes in $O_{\text{a}}$ provide two types of information: a) the *direction* in which information arrives to the active observers, which uniquely determines a subset $\mathcal{T}_{\text{a}} \subseteq \mathcal{T}$ of regular nodes (called *active subtree*, Fig. 2a); and b) the *timing* at which the information arrives to the active observers, denoted by $\{t_k\}_{k=1}^{K_{\text{a}}}$, which is used to localize the source within the set $\mathcal{T}_{\text{a}}$. It is also convenient to label the edges of $\mathcal{T}_{\text{a}}$ as $E(\mathcal{T}_{\text{a}}) = \{1, 2, \ldots, E_{\text{a}}\}$, so that the propagation delay associated with edge $i \in E$ is denoted by the RV $\theta_i$ (Fig. 2a).

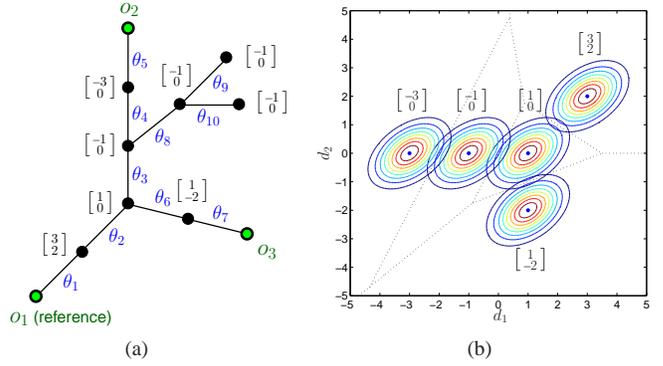

Figure 2. (a) Active tree $\mathcal{T}_{\text{a}}$. The vector next to each candidate source $s$ is the normalized deterministic delay $\tilde{\boldsymbol{\mu}}_s \triangleq \boldsymbol{\mu}_s / \mu$. The normalized delay covariance for this tree is $\tilde{\boldsymbol{\Lambda}} \triangleq \boldsymbol{\Lambda}/\sigma^2 = [5, 2; 2, 4]$. (b) Equiprobability contours of the PDFs $\mathbf{P}(\mathbf{d}|s^* = s)$ for all $s \in \mathcal{T}_{\text{a}}$, and the corresponding decision regions. For a given observation $\mathbf{d}$, the optimal estimator chooses the source $s$ that maximizes $\mathbf{P}(\mathbf{d}|s^* = s)$.

We consider that the propagation delays associated with the edges of $\mathcal{T}$ are independent identically distributed (i.i.d.) RVs with Gaussian distribution $\mathcal{N}(\mu, \sigma^2)$, where the mean $\mu$ and variance $\sigma^2$ are known [15]. With these definitions, we have the following result.

*Proposition 1 (Optimal Estimation in General Trees):* For a general propagation tree $\mathcal{T}$, the optimal estimator is given by

$$
\hat{s} = \operatorname*{argmax}_{s \in \mathcal{T}_{\text{a}}} \boldsymbol{\mu}_s^{\mathrm{T}} \boldsymbol{\Lambda}^{-1} \left( \mathbf{d} - \frac{1}{2} \boldsymbol{\mu}_s \right) \quad (2)
$$

where $\mathbf{d}$ is the *observed delay*, $\boldsymbol{\mu}_s$ is the *deterministic delay*, and $\boldsymbol{\Lambda}$ is the *delay covariance*, given by

$$
[\mathbf{d}]_k = t_{k+1} - t_k, \quad (3)
$$

$$
[\boldsymbol{\mu}_s]_k = \mu \cdot (|\mathcal{P}(s, o_{k+1})| - |\mathcal{P}(s, o_1)|), \quad (4)
$$

$$
[\boldsymbol{\Lambda}]_{k,i} = \sigma^2 \cdot \begin{cases} |\mathcal{P}(o_1, o_{k+1})|, & k = i, \\ |\mathcal{P}(o_1, o_{k+1}) \cap \mathcal{P}(o_1, o_{i+1})|, & k \neq i, \end{cases} \quad (5)
$$

for $k, i = 1, \ldots, K_{\text{a}} - 1$, with $|\mathcal{P}(u, v)|$ denoting the number of edges (length) of the path connecting vertices $u$ and $v$.

Intuitively, $\boldsymbol{\mu}_s$ and $\boldsymbol{\Lambda}$ represent, respectively, the *mean* and *covariance* of the observed delay $\mathbf{d}$ (a random vector), when node $s$ is chosen as the source (see Figure 2 for visual interpretation). The full proof of Proposition 1 is given in [16, sec. S1].

Proposition 1 essentially reduces the estimation formula in (1) to a tractable expression whose parameters can be simply obtained from *path lengths* in the tree $\mathcal{T}$. Furthermore, it is easy to show that the complexity of (2)-(5) scales as $O(N)$ with the number of nodes $N$ in the tree [16, sec. S2]. In practice, the Gaussian condition for the propagation delays can often be relaxed to non-Gaussian scenarios. The estimator in Proposition 1 can be shown to be near-optimal (see [16, sec. S3] for a concrete example), as long as the observers are *sparse*—which is often verified in practice—and the propagation delays have finite moments. The sparsity implies that the distance between observers is large, and so is the number of

RVs of the sum

$$d_k = t_{k+1} - t_1 = \sum_{i \in \mathcal{P}(s^*, o_{k+1})} \theta_i - \sum_{i \in \mathcal{P}(s^*, o_1)} \theta_i.$$

Then, the observer delay vector $\mathbf{d}$ can be closely approximated by a Gaussian random vector, due to the central limit theorem.

We now consider the most general case of source estimation on an arbitrary graph $\mathcal{G}$. When the information is diffused on the network, there is a *tree* corresponding to the first time each node gets informed, which spans all nodes in $\mathcal{G}$. Since the number of spanning trees can be exponentially large, we introduce an approximation by assuming that the actual diffusion tree is a *breadth-first search* (BFS) tree. This corresponds to assuming that the information travels from the source to each observer along a minimum-length path, which is intuitively satisfying. The resulting estimator can be written as

$$\hat{s} = \underset{s \in \mathcal{G}}{\operatorname{argmax}} \, \mathcal{S}(s, \mathbf{d}, \mathcal{T}_{\text{bfs},s}), \tag{6}$$

where $\mathcal{S} = \boldsymbol{\mu}_s^{\mathrm{T}} \boldsymbol{\Lambda}_s^{-1} \left(\mathbf{d} - \frac{1}{2}\boldsymbol{\mu}_s\right)$, with parameters $\boldsymbol{\mu}_s$ and $\boldsymbol{\Lambda}_s$ computed with respect to the BFS tree $\mathcal{T}_{\text{bfs},s}$ rooted at $s$. It can easily shown that the complexity of (6) scales subexponentially with $N$, as $O(N^3)$ [16, sec. S2].

We now turn our attention to the localization performance and its dependence on: i) the structure of the network, ii) the density and placement of the observers, and iii) the observation of multiple information cascades. We first apply the proposed estimator to various synthetic networks, shown in Table 1. Clearly, the estimator performs the best in scale-free networks (such as the Barabási-Albert [17][18] and the Apollonian models [19–21])—in some cases requiring as few as 4% of observers to achieve a localization probability of 90%. This is because scale-free networks exhibit "hubs" with large degrees, which can be picked as observers and are able to receive a large amount of information about the source. If the network is not scale-free (such as the Erdös-Rényi model), or the observers are placed uniformly at random, then more observers are necessary to achieve the same localization performance.

So far we assumed that the source of information transmits only one message. However, in many scenarios, the source emits different messages over time, which diffuse independently over the network. These *information cascades* can be gathered and exploited by the observers, as revealed by the following proposition.

*Proposition 2 (Effect of Multiple Cascades):* If the source $s^*$ transmits $C$ independent cascades of information on a tree $\mathcal{T}$, then the probability of correct localization $P_{\text{loc}}$ achieved by the optimal estimator is given by

$$P_{\text{loc}} = P_{\text{max}} - O\left(e^{-aC}\right),$$

where $P_{\text{max}}$ is the maximum probability of localization achieved under *deterministic* propagation, and $a$ is a constant.

The full proof is given in [16, sec. S4]. The proposition shows that as the observers collect more information from successive cascades, they can average out the variance associated with random propagation, and approach the localization performance of the deterministic scenario ($P_{\text{max}}$) at a rate that is at least *exponential*. We can think of such phenomenon as a *time-resolution tradeoff*: the observers can achieve higher accuracy of localization by waiting for a longer time, over which they can observe more cascades.

Table 1. Percentage $K/N$ of observers necessary to achieve $P_{\text{loc}} = 90\%$, for different networks and observer placements. The "high-degree" placement picks the highest-degree nodes as observers, while the "random" placement picks the observers randomly. We consider $N = 100$ nodes, and propagation ratio $\mu/\sigma = 4$.

| Network | Observer Placement | |
|---|---|---|
| | High-degree | Random |
| Apollonian | 4% | 25% |
| Barabási-Albert | 18% | 41% |
| Erdös-Rényi ($Np = 0.5$) | 34% | 49% |
| Erdös-Rényi ($Np = 2$) | 32% | 44% |

Lastly, we test the effectiveness of the proposed algorithm with real, measured data. We consider the well-documented case of cholera outbreak that occurred in the KwaZulu-Natal province, South Africa, in 2000 (Fig. 3a). The epidemic was caused by a strain of *Vibrio cholerae*, which colonizes the human intestine and is transmitted through contamination of aquatic environments. The dataset was provided by the KwaZulu-Natal Health Department, and consists of each single cholera case, specified by the date and health subdistrict where it occurred. To perform source localization, we consider a network model of the basin (Fig. 3b) developed in [10]. The nodes represent human communities and associated water reservoirs, in which the disease can be diffused and grow. The edges of the graph represent hydrological links between the communities. The propagation parameters for this bacteria were obtained from past epidemics [16, sec. S5][22]. Source localization is performed by monitoring the daily cholera cases reported in $K$ communities (the observers). These are selected uniformly at random, due to the lack of a priori information about the source location. Table 3c shows that by monitoring only 20% of the communities, we achieve an average error of *less than 4 hops* between the estimated source and the first infected community. This small distance error may enable a faster emergency response from the authorities in order to contain an outbreak.

To conclude, the results in this paper suggest that a sparse deployment of observers may provide an effective alternative to the individual monitoring (either human or automatic) of all nodes in a network. However, several challenges remain. First, in some scenarios, it may be difficult to exactly determine the underlying graph on which diffusion occurs. In a cholera outbreak, for example, the diffusion of the bacteria is also influenced by the long-range movement of infected individuals, in addition to the basic hydrological transport. Since this mobility network cannot be reliably measured, further study is needed to determine the robustness of our framework to inaccuracies in the underlying graph. Second, the choice of observers in the network strongly affects the performance of the proposed algorithm. Optimal strategies for observer placement need to be further investigated. Nevertheless, our results indicate that source localization in large networks—

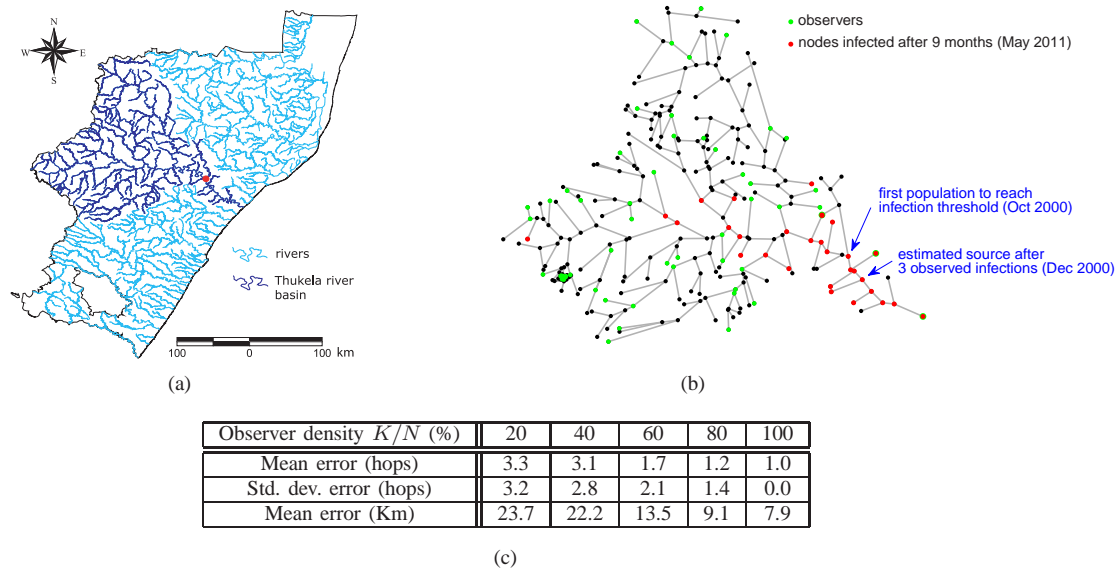

Figure 3. Locating the source of a cholera outbreak in the KwaZulu-Natal province in South Africa. (a) Hydrographic map of the KwaZulu-Natal province. The red dot corresponds to the location of the first reported cases of cholera. (b) Graphical model of the Thukela river basin. Nodes represent small communities and associated water reservoirs, in which the disease can be diffused and grow. The edges reflect the transport of cholera between neighboring communities, due to hydrological flow and human mobility. To localize the source of the outbreak, we monitor 20% of the communities, selected at random (in green). (c) Average distance between the estimated source and the first infected community, versus the observer density $K/N$. With 20% of observers, we achieve an average error of less than 4 hops. Note that the first infected community is not necessarily the source of the outbreak, due to the delay between the infection and the actual reporting of the disease.

| Observer density $K/N$ (%) | 20 | 40 | 60 | 80 | 100 |
|---|---|---|---|---|---|
| Mean error (hops) | 3.3 | 3.1 | 1.7 | 1.2 | 1.0 |
| Std. dev. error (hops) | 3.2 | 2.8 | 2.1 | 1.4 | 0.0 |
| Mean error (Km) | 23.7 | 22.2 | 13.5 | 9.1 | 7.9 |

(c)

a seemingly impossible task with only a few sensors—is indeed feasible, both in terms of localization accuracy and computational cost.

We thank Andrea Rinaldo, Enrico Bertuzzo, and Lorenzo Mari at EPFL for providing the data of the cholera outbreak. We also thank Paolo De Los Rios for helpful comments. This work was supported by the ERC Advanced Grant – Support for Frontier Research – SPARSAM Nr: 247006.